\documentclass[pre,twocolumn,showpacs,preprintnumbers,amsmath,amssymb,superscriptaddress]{revtex4}

\usepackage{graphicx}
\usepackage{dcolumn}

\newcommand{\xbo}{{\mathbf{x}}}

\newcommand{\qbo}{{\mathbf{q}}}
\newcommand{\nbo}{{\mathbf{n}}}

\newcommand{\beq}{\begin{equation}}
\newcommand{\eeq}{\end{equation}}

\newcommand{\Ocal}{{\cal O}}

\newcommand\mean[1]{\langle#1\rangle}

\begin{document}


\title{Defect formation in the Swift-Hohenberg equation}
\author{Tobias Galla}
\email{galla@thphys.ox.ac.uk}
\affiliation{Theoretical Physics, University of Oxford, 1 Keble Road
OX1 3NP, United Kingdom}
\author{Esteban Moro}
\email{emoro@math.uc3m.es}
\affiliation{Theoretical Physics, University of Oxford, 1 Keble Road
OX1 3NP, United Kingdom}
\affiliation{Departamento de Matem\'aticas \& GISC, Universidad
Carlos III de Madrid, Avenida Universidad 30, 28911 Legan\'{e}s, Spain}

\date{\today}

\begin{abstract}
  We study numerically and analytically the dynamics of defect
  formation during a finite-time quench of the two dimensional
  Swift-Hohenberg (SH) model of Rayleigh-B\'enard convection. We find
  that the Kibble-Zurek picture of defect formation can be applied to
  describe the density of defects produced during the quench.
  Our study reveals the relevance of two
  factors: the effect of local variations of the striped patterns
  within defect-free domains and the presence of both point-like and
  extended defects. Taking into account these two aspects we are able
  to identify the characteristic length scale selected during the
  quench and to relate it to the density of defects. We discuss
  possible consequences of our study for the analysis of the
  coarsening process of the SH model.
\end{abstract}
\maketitle


The formation of topological defects in symmetry-breaking phase
transitions is a very generic phenomenon in physics, and can be
studied analytically and experimentally in different condensed matter
systems \cite{chaikin,hohenberg}. An example is the onset and
formation of stripe patterns in Rayleigh-B\'enard convection
\cite{hohenberg,bodenschatz}. In this paper, we focus on the
Swift-Hohenberg (SH) model \cite{sh} for this process. Once above the
convective threshold, the system develops a labyrinthine morphology,
consisting of domains of stripes which are oriented along
arbitrary directions \cite{seul}. Between those domains, the system
displays several types of topological defects such as grain
boundaries, disclinations and dislocations. This structure orders with
time basically by grain boundary relaxation and defect annihilation.
Similar to other models, in which this coarsening
process is self-similar \cite{bray}, any linear scale of the structure
is expected to grow as a power law in time $\xi \sim t^{1/z}$.
However, simulations of sudden quenches of the SH equation
\cite{elder,cross,goldenfeld,bray2,boyer,mazenko} have revealed that
the observed exponent $z$ is sensitive to non-universal model features
such as the quench depth or noise strength. Moreover, different
definitions of the length scale have led to different exponents with
values reported in the interval $2 \leq z \leq 5$. This apparent
absence of self-similarity in the coarsening is also found in related
experiments of electroconvection \cite{purvis} and diblock copolymers
\cite{harrison}. Multiscaling \cite{bray2,purvis} and/or defect
pinning \cite{boyer} have been proposed as being responsible for the
scattered value of $z$, but no general picture has been reached so far
about the true nature of the coarsening process.

In this paper we consider a complementary, but related aspect of the
non-equilibrium dynamics of the SH equation.  Namely, we are
interested in the formation of defects in a finite-time quench
(annealing). Interestingly, some features of finite-time quenches were
considered in some early works comparing the SH model with
Rayleigh-B\'enard experiments \cite{ahlers}. Specifically, we study
situations in which the control parameter, the reduced Rayleigh number
$\varepsilon \equiv (R-R_c)/R_c$, is swept smoothly over the
bifurcation point, $\varepsilon=0$. Our study is thus related to
earlier papers on defect formation in nonequilibrium second-order
phase transitions \cite{zurek,zurek1,Lythe}. The theoretical
picture which is believed to be applicable in this case is the
Kibble-Zurek mechanism \cite{zurek}: because of critical slowing down
close to the transition point, the system can not follow the external
change of the critical parameter and the dynamics are suspended in an
interval $[-\hat \varepsilon,\hat \varepsilon]$ around the transition.
After this suspension a characteristic length $\xi$ is selected and is
found to scale with the rate of increase, $\mu$, of the critical
parameter like $\xi \sim \mu^{-\gamma}$. In particular, mean-field
theory predicts $\hat \varepsilon\sim\mu^{1/2}$ and
$\xi\sim\mu^{-1/4}$ for models of the Ginzburg-Landau type with $O(N)$
symmetry \cite{zurek}.  This length $\xi$ sets an initial density of
defects, $\rho$, directly after the quench. Within a Gaussian
approximation one expects $\rho \sim \xi^{-2}$ for point-like defects
and $\rho \sim \xi^{-1}$ for line defects in two dimensions
\cite{liumazenko}. Thus, the Kibble-Zurek mechanism predicts that the
density of defects observed directly after the quench scales with
$\mu$ as well, as confirmed in various models \cite{zurek1,Lythe,casado}.

The purpose of this paper is to confirm the validity of the
Kibble-Zurek scenario for the SH model. To this end, we identify the
characteristic length $\xi$ selected during the annealing process and
study its relation to the density of defects.  We model the annealing
protocol of the Rayleigh-B\'enard system by the SH equation in two
dimensions and in dimensionless variables \cite{sh,hohenberg}
\begin{equation}\label{sh}
\partial_t\phi  = \varepsilon(t)\phi-(q_0^2+\nabla^2)^2\phi -\phi^3 + \eta.
\end{equation}
where $\varepsilon(t) = \mu t$. This corresponds to an experimental
situation where the temperature difference between the upper and the
lower plate of the convection cell is increased linearly in time. The
order parameter field $\phi(\xbo,t)$ is related to the vertical fluid
velocity. The last term in Eq.\ (\ref{sh}) is a stochastic forcing
term, with $\mean{\eta(\xbo,t)\eta(\xbo^\prime,t^\prime)}= 2 F
\delta(\xbo-\xbo^\prime)\delta(t-t^\prime)$, where the noise strength
$F = 5\times 10^{-17}$ is compatible with typical experimental values
\cite{sh,ahlers} . Our simulations are performed on a square lattice
of $512\times 512$ nodes with lattice spacing $\Delta x = \pi/4$,
corresponding to 8 lattice points per ideal wavelength ($q_0=1$).
Initial conditions are $\phi(\xbo,t_0)=0$ at the instant $t_0$ given
by $\varepsilon(t_0)=-1/2$.

\begin{figure}
\begin{center}
\includegraphics[width=2.2in,angle=-90]{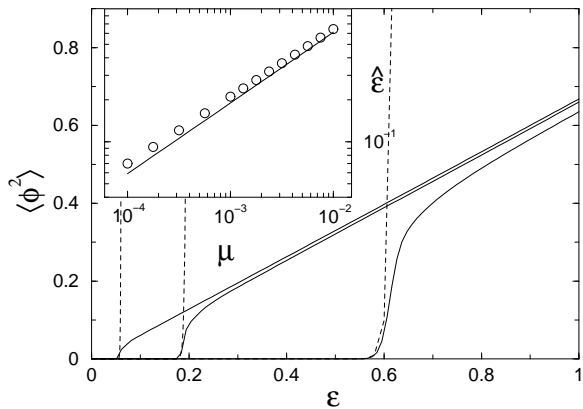}
\caption[phi2]{\label{phisq}
  Variance of the order parameter, $\mean{\phi^2}$, obtained from the
  simulations (solid lines) compared with the linear approximation
  given by Eq.\ (\ref{phisqana}) (dashed lines) with $\mu=10^{-4}$,
  $\mu=10^{-3}$ and $\mu=10^{-2}$ from left to right. Inset: numerical
  values of $\hat \varepsilon$ as a function of $\mu$ obtained
  implicitly from $\mean{\phi^2} = \delta \hat \varepsilon$ with
  $\delta = 1/2$. Solid line is the prediction of Eq.\ (\ref{ghat}).}
\end{center}
\end{figure}

\begin{figure}
\begin{center}
\includegraphics[width=8.5cm]{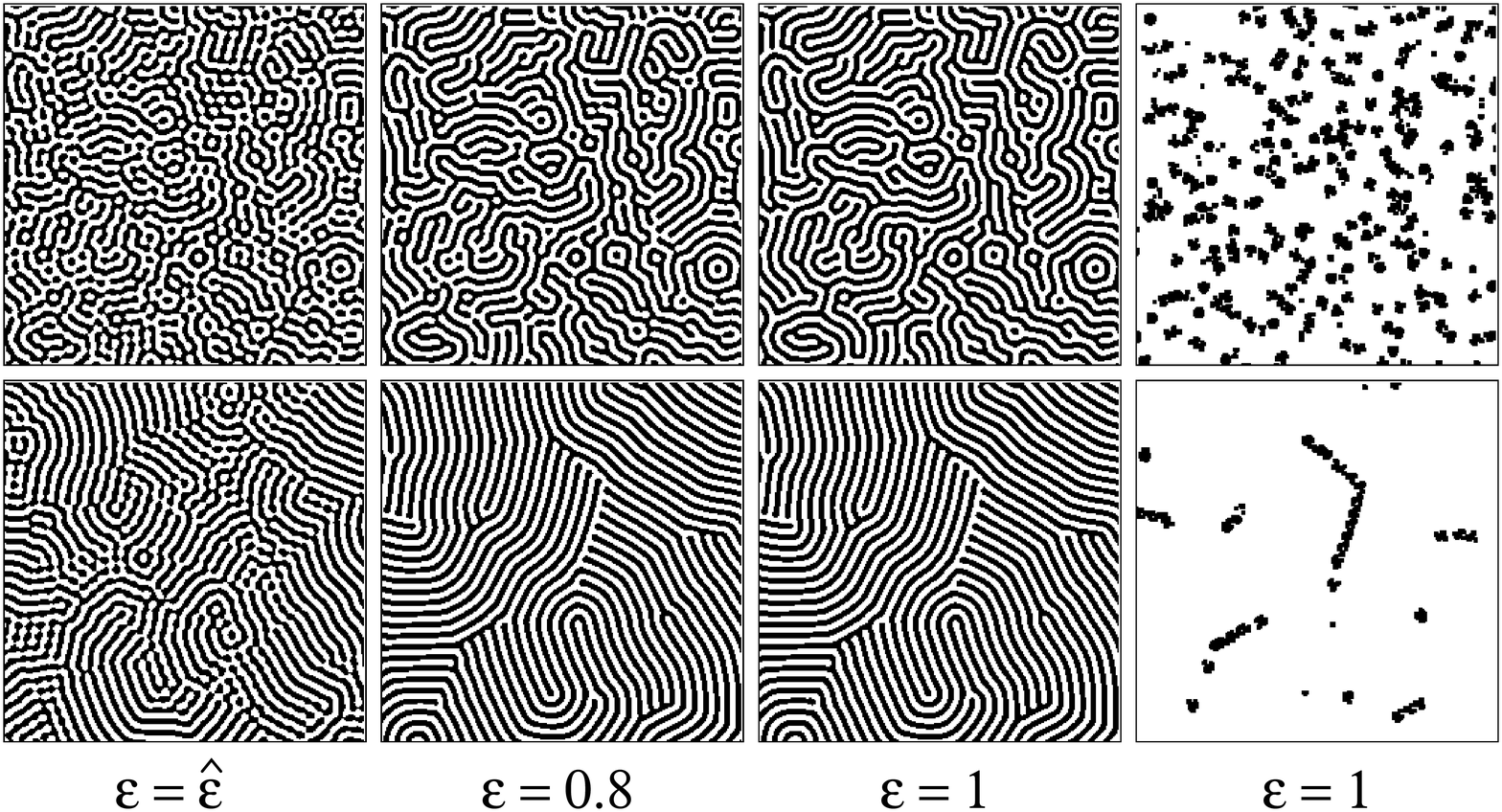}
\caption[signal]{\label{signal} 
  Typical configurations of the order parameter field for different
  values of $\varepsilon$. We depict an area of $256\times 256$
  lattice points. Upper row: $\mu=10^{-2}$, lower row $\mu=10^{-4}$.
  Black (white) points correspond to $\phi>0$ ($\phi<0$).  Rightmost
  panels display the defect structures obtained from the amplitude
  signal ($\vartheta=0.25$, see text).}
\end{center}
\end{figure}

As the critical parameter $\varepsilon$ is increased in time the
magnitude of the order parameter remains close to $\phi=0$ until well
after the onset of the instability at $\varepsilon=0$ (see Fig.\ 
\ref{phisq}). At some later instant $\varepsilon=\hat \varepsilon>0$
the field abruptly jumps towards its symmetry-broken quasi-equilibrium
at $\mean{\phi^2}=2\varepsilon(t)/3$ and spatially periodic
modulations of the signal are created, see Fig.\ \ref{signal}. We
identify the instant $\hat\varepsilon$ at which dynamics are resumed
as the time when $\mean{\phi^2} = \delta \hat \varepsilon$
\cite{Lythe} [with $\delta=\Ocal(1)$, meaning that the first and third
term on the right-hand side of Eq.\ (\ref{sh}) are of equal
importance]. Results are presented in Fig.\ \ref{phisq} where we find
$\hat \varepsilon \sim \mu^{0.48\pm 0.01}$, which compares well with
the Kibble-Zurek predicted scaling $\hat \varepsilon\sim \mu^{1/2}$.
It is interesting to note that the same scaling was found in ramp
experiments of Rayleigh-B\'enard convection \cite{ahlers}, where the
temperature difference between the upper and lower plate of the
convection cell was increased linearly in time. This setup is similar
to the annealing protocol in our simulations suggesting that the
scaling found in those experiments can be explained within the
Kibble-Zurek scenario as well.

\begin{figure}
\includegraphics[height=8.0cm,angle=270]{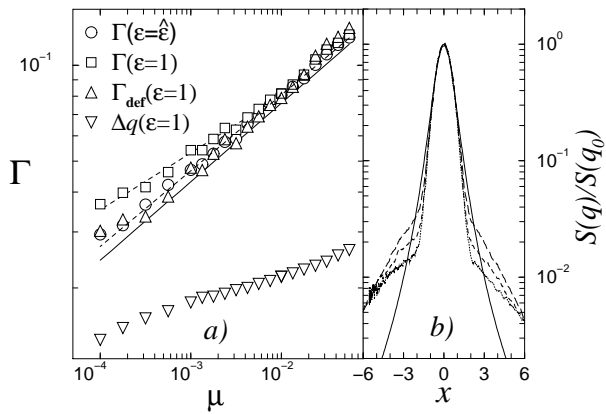}
\caption[collapse]{\label{sofq} a) Width of the structure factor,
  $\Gamma$, after the quench compared with the local variations of the
  wave-number $\Delta q$. Solid line is the approximation of Eq.\ 
  (\ref{width}), dashed lines are fits to power laws
  $\Gamma(\varepsilon=\hat\varepsilon) \sim \mu ^{0.24\pm 0.01} $and
  $\Gamma(\varepsilon=1) \sim \mu^{0.18\pm0.01}$.  b) Collapsing plots
  of $S(q)$ as a function of the rescaled variable
  $x=(q^2-q_0^2)/(q_0\Gamma)$ at $\varepsilon=\hat \varepsilon$ for
  $\mu=10^{-2}$ (dotted line), $\mu=10^{-3}$ (dashed line) and
  $\mu=10^{-4}$ (long dashed line); the solid line is a squared
  Lorentzian curve $1/(1+x^2)^2$.}
\end{figure}

Since the order parameter remains small up to $\varepsilon = \hat
\varepsilon$, the observed scaling can be obtained from a simple linear approximation of Eq.\ (\ref{sh}). Fourier transforming the
linearized SH-equation yields
\begin{equation}\label{shfourier}
\partial_t\hat\phi_{\qbo}(t)= \left[\mu t
-(q_0^2-\qbo^2)^2\right]\hat\phi_\qbo(t)+\hat\eta_\qbo(t).
\end{equation}
where $\hat \phi_\qbo(t)$ is the Fourier mode of the order parameter
field with wave vector $\qbo$.  Thus, the structure factor $S(q,t)
\equiv \mean{|\hat\phi_\qbo(t)|^2}$ is given by
\begin{equation}
S(q,t)= 2 F \mathrm{e}^{2\omega_q(t)}\int_{t_0}^t e^{-2\omega_q(s)}ds
\end{equation}
where $\omega_q(t) \equiv \frac12
\mu(t^2-t_0^2)-(q_0^2-q^2)^2(t-t_0)$. The structure factor $S(q,t)$ is
peaked around $q \simeq q_0$ (see Fig.\ \ref{sofq}b) and the time
dependent width of this peak, $\Gamma(t)$, is obtained from $S[q_0 \pm
\frac12\Gamma(t),t] \equiv \frac12 S(q_0,t)$. We find that $\Gamma(t)$
satisfies the implicit equation
\begin{equation}\label{Gamma}
\Gamma(t)=\left(\frac{\ln(2 \alpha)}{2q_0^2 t}\right)^{1/2},
\end{equation}
where $\alpha = 1 + \Ocal(\Gamma^2)$. Assuming that $S(q,t)$ is
sharply peaked around $q_0$ (i.e.\ $\Gamma \ll 1$) and that it can be
approximated by a squared Lorentzian around $q_0$
\cite{cross,goldenfeld} (see Fig.\ \ref{sofq}b) we write
\begin{equation}\label{phisqana}
\mean{\phi^2}= \sum_{\qbo}
S(q,t) \simeq \frac{q_0 S(q_0,t)}{8(\sqrt{2}-1)^{1/2}}\;\Gamma(t).
\end{equation}
Combining this result with the implicit equation $\mean{\phi^2}=\delta
\hat \varepsilon$, we find that $\hat \varepsilon$ satisfies
\begin{equation}\label{ghat}
\hat \varepsilon^2 \simeq \mu \ln[(2F)^{-1}\delta q_0 C_0 \hat \varepsilon^{3/2}]
\end{equation}
where $C_0 \simeq 4.93$ is a numerical constant. Thus, the linear
approximation leads, up to logarithmic corrections, to the scaling
behavior $\hat \varepsilon\sim\mu^{1/2}$ predicted by the
Kibble-Zurek scenario and confirmed by our simulations (see Fig.\ 
\ref{phisq}).

We now proceed to identify the typical length scale selected by the
dynamics during the quench. A canonical measure for this
length is the width of the structure factor
\cite{cross,elder,goldenfeld,bray2,boyer,mazenko}: in a defect-free domain the
order parameter field takes the form
\begin{equation}
\label{smectic}
\phi(\xbo,t) \simeq A(\xbo,t)
\cos(\qbo(\xbo,t) \cdot \xbo) {\mathrm{ + higher\ harmonics}}
\end{equation}
where $A(\xbo,t)$ and $\qbo(\xbo,t)$ vary slowly within the domain.
Thus, in the absence of any defects one expects the structure factor
$S(q,t)= \mean{|\hat \phi_\qbo(t)|^2}$ of an infinite system to be a
delta function centered around $q = q_0$. When defects are present,
the structure factor broadens around $q_0$ and the width $\Gamma$ at
half-height can be used as a proxy for the length, i.e.\ we define
$\xi\equiv\Gamma^{-1}$.  Fig.\ \ref{sofq}a shows the data for $\Gamma$
at $\varepsilon = \hat \varepsilon$ and $\varepsilon = 1$ numerically
obtained from a circular average of the structure factor.  We find
that $S(q,t)$ has a scaling form $S(q,t)/S(q_0,t) \simeq
f[(q^2-q_0^2)/(q_0\Gamma)]$ (see Fig.\ \ref{sofq}b) at least
for a given (scaling) interval around $q=q_0\simeq 1$.  Results are
thus independent of the threshold (half-height) used to obtain
$\Gamma$, provided the analysis is performed within this scaling
region. At $\varepsilon=\hat\varepsilon$ we find $\xi \sim \mu^{-0.24
  \pm 0.01}$ which again confirms the Kibble-Zurek prediction $\xi\sim
\mu^{-1/4}$.  Note that this scaling can be recovered within the
linear approximation by combining Eqs.\ (\ref{Gamma}) and (\ref{ghat})
to obtain
\begin{equation}\label{width}
\Gamma(\hat\varepsilon) \simeq \mu^{1/4}\sqrt{\frac{\ln 2}{2q_0^2}}\;
[\ln(\delta q_0 C_0 \hat \varepsilon^{3/2} (2F)^{-1})]^{-1/4}
\end{equation}
which agrees with the Kibble-Zurek result up to logarithmic
corrections and compares quantitatively well with the numerical data
(see Fig.\ \ref{sofq}a).

We find, however, that right after the jump the absolute value of
$\xi$ decreases and then rapidly saturates to a value smaller than the
one at the jump. The data at any later instant (e.g.\ $\varepsilon=1$)
deviates from the Kibble-Zurek prediction in the regime of small
$\mu$, in fact in the interval $10^{-4}\leq\mu\leq 10^{-3}$ it can be fitted
to $\xi \sim \Gamma^{-1} \sim \mu^{-0.18\pm 0.01}$.  To
understand this apparent inconsistency, we first note that coarsening
of the defect structure cannot be responsible for this effect, since
$\xi(\varepsilon=\hat \varepsilon) > \xi(\varepsilon=1)$.  Instead, a
possible explanation is given in \cite{egolf,purvis}: the broadening
of the spectrum and hence the observed value of $\Gamma$ is not only
related to the spatial distribution of defects, but also to variations
of the local wavenumber $q(\xbo,t)$ within the defect-free domains.
Using the numerical procedure suggested in \cite{egolf} we find that
although the distribution of local wave-numbers $q(\xbo,t)$ in
(\ref{smectic}) is centered around the expected value $q(\xbo,t) =
q_0$, it exhibits a finite width indicating variations in the actual
values of $q(\xbo,t)$. To quantify this effect we measure the RMS of
those fluctuations, $\Delta q \equiv \mean{[q(\xbo,t)-q_0]^2}^{1/2}$,
where the average extends over defect-free domains only. We find that
$\Delta q$ is a significant fraction of $\Gamma$ for small values of
$\mu$, where $\Gamma$ is small (see Fig.\ \ref{sofq}a). Although in
principle the observed value of $\Gamma$ can be a complicated
convolution of both the local variations of the value of $q(\xbo,t)$
and the presence of defects, we have tried a simple linear Ansatz to
disentangle both contributions, i.e.\ $\Gamma =
\Gamma_{\mathrm{def}}+\Delta q$, where we assume that
$\Gamma_{\mathrm{def}}$ is directly related to the spatial
distribution of defects. In fact, the data for $\Gamma_{\mathrm{def}}$
at $\varepsilon=1$ obtained using this simple Ansatz can be fitted to
a power law $\Gamma_{\mathrm{def}}\sim\mu^{-0.24\pm 0.01}$ and
rescaled by a constant factor to collapse well with $\Gamma$ taken at
$\varepsilon=\hat\varepsilon$ as shown in Fig.\ \ref{sofq}a. This
indicates that the Kibble-Zurek scaling $\xi \sim \mu^{-1/4}$ might
still hold at $\varepsilon = 1$, although it appears to be masked by
local variations of the wavenumber \cite{jump}.

An independent way of checking the Kibble-Zurek predictions is to
measure the density of defects, $\rho$, created during the quench.
Several methods have been devised to identify defects numerically
\cite{boyer,bray2,cross,goldenfeld,mazenko}. Here we present results
for two independent measures of the number of defects: the first one
is based on the local amplitude of the order parameter field
\cite{goldenfeld}, while the second one relies on the local curvature
of the observed stripes \cite{boyer}. Far away from any defect the
order parameter field $\phi(\xbo,t)$ is of the sinusoidal form
(\ref{smectic}), so that the local amplitude can be numerically
estimated as $A^2(\xbo,t) \simeq
\phi(\xbo,t)^2+(\nabla\phi(\xbo,t))^2/q_0^2$. In equilibrium, it is
found that $A^2(\xbo,t) \simeq A_0^2 \equiv 4\varepsilon/3$ to lowest
order in $\varepsilon$ for a set of parallel stripes \cite{hohenberg}.
The density of defects, $\rho_a$, can thus be calculated using the
filter $|[A^2(\xbo,t)-A_0^2]/A_0^2| > \vartheta$, where $\vartheta$ is
some threshold. The second method consists in numerically extracting
the probability distribution $P(\kappa,t)$ of the local curvature
$\kappa(\xbo,t)=|\nabla\cdot\nbo(\xbo,t)|$ of the stripe patterns
\cite{boyer}. Here $\nbo=\nabla\phi/|\nabla\phi|$ is the unit vector
normal to the lines of constant $\phi$. For a defect-free pattern, the
curvature is zero, so again defect points can be identified as those
with $\kappa(\xbo,t) > \Theta$, where $\Theta$ is a given threshold.
Thus the density of defects is $\rho_c = \int_\Theta^\infty
P(\kappa,t) d\kappa$. In Fig.\ \ref{signal} we display typical
configurations of the SH signal and the corresponding defect
structures obtained using the first filter, demonstrating that this
method can detect the defect structure efficiently. The thresholds
$\vartheta$ and $\Theta$ are carefully chosen to be within the finite
interval in which our results for $\rho_a$ and $\rho_c$ are (up to a
constant prefactor) independent of the particular choice of
$\vartheta$ and $\Theta$.

\begin{figure}
\includegraphics[width=5.8cm,angle=270]{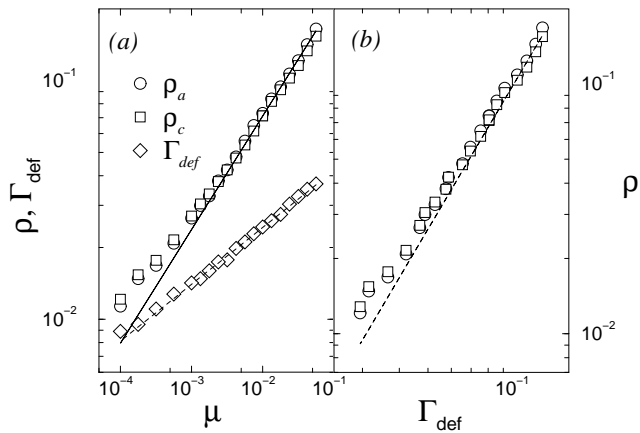}
\caption[rhod]{\label{rhod} (a) Density of defects $\rho_a$ and
$\rho_c$ (rescaled to collapse with $\rho_a$) obtained at
$\varepsilon=1$ (the data for $\mu > 10^{-2}$ is taken at $\varepsilon
= 2$), and $\Gamma_{\mathrm{def}}=\Gamma(\varepsilon=\hat\varepsilon)$
(in arbitrary units) as a function of $\mu$,. Solid line corresponds
to $\rho \sim \mu^{0.48}$ and dashed line is $\Gamma_{\mathrm{def}}
\sim \mu^{0.24}$. (b) Density of defects as a function of
$\Gamma_{\mathrm{def}}$. Dashed line is $\rho \sim
\Gamma_{\mathrm{def}}^2$.}
\end{figure}

As shown in Fig.\ \ref{rhod} the numerical values for $\rho_c$ can be
rescaled to agree very well with the density of defects obtained from
the amplitude signal, $\rho_a$, so that (up to a constant factor) both
methods independently yield the same density of defects. For
relatively large values of $\mu$, we find that the density scales like
$\rho \sim \mu^{1/2}$ and also that $\rho \sim
\Gamma_{\mathrm{def}}^2$, as shown in Fig.\ \ref{rhod}. Consequently,
if $\Gamma_{\mathrm{def}}$ is related to the density of defects, those
should be point-like for quick quenches. A simple inspection of the
configurations after the jump confirms this picture, see Fig.\ 
\ref{signal}: we observe that for small values of $\mu$ most of the
defects are point-like, while only a minor fraction is spatially
extended. Only at small values of the sweep rate does the density of
defects deviate from the $\rho \sim \Gamma_{\mathrm{def}}^2 \sim
\mu^{1/2}$ scaling, and we observe an increasingly significant amount
of extended defects. If extended defects dominate over point-like
defects, a crossover to $\rho \sim \Gamma_{\mathrm{def}} \sim
\mu^{1/4}$ should be expected for small $\mu$. Although a departure
from the $\rho \sim \mu^{1/2}$ behaviour is evident for slow quenches,
our computational facilities at present do not allow us to give an
accurate confirmation of this possible crossover.

In conclusion, we have confirmed the validity of the Kibble-Zurek
picture for the creation of defects in the annealing of the SH
equation. Specifically, we have identified a length scale $\xi \sim
\Gamma_{\mathrm{def}}^{-1}$ directly related to the density of
defects, and found that it scales with the sweep rate as $\xi \sim
\mu^{-1/4}$. For large values of $\mu$, where most of the created
defects are point-like, we expect $\rho \sim \xi^{-2}$, a result which
is observed in our simulations.  However for slow quenches, where an
increasing fraction of extended defects is produced, we expect a
deviation from this scaling law and a possible crossover to $\rho \sim
\xi^{-1}$. We hope this work will stimulate other simulations of the
SH model in order to confirm this picture.

Finally, we note that in studies of the coarsening in the SH model
\cite{boyer,bray}, it is observed that at large times defects are
extended and that thus $\rho \sim \xi^{-1}$. Our study suggests that
the presence of point-like defects might be relevant, and thus some
care has to be taken in the SH model when comparing the density of
defects with the characteristic length observed in the system. We hope
that taking into account this aspect together with the identification
of the length scale associated with the density of defects as
suggested in this paper might be useful to shed some light on the
question of self-similarity in the coarsening process after an
sudden quench of the SH system \cite{ournew}.

We thank G. Lythe for participation in the early stages of this work
and D.\ Boyer and J.\ Vi\~nals for discussions.  Support is
acknowledged from EPSRC (UK) Grant GR/M04426 and Studentship 00309273,
and EU fellowship HPMF-CT-2000-0487. TG acknowledges the support of
the Rhodes Trust and Balliol College Oxford.


\begin{thebibliography}{99}
\bibitem{jump} Note that directly at the jump striped patterns are not
  yet fully developed making the definition of local wavenumber,
  curvature and/or defects meaningless.
\end{thebibliography}
\end{document}